\documentclass[aps,pra,twocolumn,10pt]{revtex4-1}
\usepackage{graphicx} 
\usepackage{amsmath,amssymb}

\begin{document}

\title{Weak measurements and nonClassical correlations}

\author{Lekshmi S.}
\affiliation{Department of Physics, Maharajas College, Ernakulam, Kerala, India 682011}
\affiliation{Department of Physics, SD College, Alappuzha, Kerala, India 688003}
\email{lekshmisdc@gmail.com}

\author{N.~Shaji}
\affiliation{Department of Physics, Maharajas College, Ernakulam, Kerala, India 682011}


\author{Anil Shaji}
\affiliation{School of Physics, Indian Institute of Science Education and Research Thiruvananthapuram, Kerala, India 695016}

\date{\today}

\begin{abstract}
We extend the definition of quantum discord as a quantifier of nonClassical correlations in a quantum state to the case where weak measurements are performed on subsystem $A$ of a bipartite system $AB$. The properties of weak discord are explored for several families of quantum states. We find that in many cases weak quantum discord is identical to normal discord and in general the values of the two are very close to each other. Weak quantum discord reduces to discord in the appropriate limits as well. We also discuss the implications of these observations on the interpretations of quantum discord. 
\end{abstract}
\maketitle

\section{\textbf{Introduction}}

In~\cite{Ollivier:PhysRevLett:2001} Ollivier and Zurek view quantum discord~\cite{Ollivier:PhysRevLett:2001,Henderson:JournalOfPhysicsAMathematicalAndGeneral:2001,Modi:ReviewsOfModernPhysics:2012} as a means of quantifying the disturbance on a system $B$ generated by a read out procedure consisting of projective measurements on an apparatus $A$ which has previously interacted with $B$ through a process of pre-measurement.  If the discord,  defined using projective measurements on $A$, is zero, then there exists a set of projective measurements on $A$ which do not alter the global state $\rho_{AB}$ of the apparatus and system  after the pre-measurement. Subsequently, the definition of quantum discord was expanded to include all possible measurements on $A$ including POVMs and discord was placed in a more general context as a measure of the nonClassical correlations that exist between two quantum systems $A$ and $B$, moving away from the `system-apparatus' picture~\cite{Datta:PhysRevLett:2008}. In constructing quantum discord and related measures~\cite{LANG:IntJQuanumInform:2011} as the difference between total correlations and classical correlations in a quantum state, wherein the classical correlations are understood as those correlations that can be `extracted' through a measurement process on either one or both of the subsystems independently, there is an implicit predisposition to do the best possible job of the measurement within the framework of the definition of each  measure. This means that despite the original motivation of Ollivier and Zurek regarding disturbance to the quantum state or parts of it, whether such disturbances occur or not is no longer a significant part of the discussion on various measures of nonClassical correlations. 

The disturbance to a measured quantum system is very much topical to the notion of weak quantum measurements introduced by Aharanov, Albert and Vaidmann ~\cite{Aharonov:PhysRevLett:1988} and later elucidated with clarity by Duck, Stevenson and Sudarsan ~\cite{Duck:PhysicalReviewD:1989}. The weak value of an observable on the measured system is defined relative to a post-selected final state of the system that, in turn, lets one pin down and choose precisely the amount of disturbance that the final state has relative to the initial state. The question of extending the notion of quantum discord for a bipartite system when weak measurements are done on one of the subsystems was considered in~\cite{singh2014quantum}, wherein the notion of super-quantum discord is introduced. We discuss discord defined with respect to weak quantum measurements in this Paper from a different point of view compared to~\cite{singh2014quantum}.  We view the increase in the numerical value of discord when weak measurements are performed on one of the subsystems as a consequence of the measurements revealing only a portion of the classical correlations contained in the bipartite systems and not as an indication of an enhanced amount of quantum correlations in the system. We are able to compute the {\em weak discord} for different families of states of two or more qubits exactly and we find that the trade off between the disturbance to the measured system due to the weak measurements and the classical correlations that are revealed by the measurement follows a common pattern across the different sets of states considered. 

In the next section we briefly review quantum discord and other nonClassical correlations and follow it by a recap of weak measurements in Sec.~\ref{sec:weakmeasurements}. The extension of quantum discord to the case where weak measurements are done on one subsystem - termed weak discord - is introduced in Sec.~\ref{sec:weakdiscord} and it is computed for various families of states. In Sec.~\ref{sec:interpretations} interpretations of quantum discord, including operational ones, are revisited in the context of weak measurements and weak quantum discord. A discussion of our results is included in Sec.~\ref{sec:conclusion}.

\section{Quantum discord with projective measurements \label{sec:discord}}

Based on entropic measures of correlations in bipartite quantum systems, several measures of nonClassical correlations can be defined by subtracting the correlations that `classical observers' interrogating one of both subsystems can detect from the total correlations between them~\cite{LANG:IntJQuanumInform:2011,Modi:ReviewsOfModernPhysics:2012}. Quantum discord~\cite{Henderson:JournalOfPhysicsAMathematicalAndGeneral:2001,Ollivier:PhysRevLett:2001}  is defined in terms of the mutual information, which is an entropic measure of correlations between two systems $A$ and $B$, in a joint state $\rho_{AB}$, defined as,  
\begin{equation}
	\label{eq:mutualI}
	I(A:B) = S\left(\rho_{A}\right) + S\left(\rho_{B}\right) - S\left(\rho_{AB}\right),
\end{equation}
where $S(\rho) = -{\rm tr}(\rho \log \rho)$ is the vonNeumann entropy of the quantum state $\rho$ and $\rho_{A,B} = {\rm tr}_{B, A}(\rho_{AB})$ are the reduced sub-system density matrices. An observer who has measured subsystem $A$ can quantify the correlations between $A$ and $B$ as
\begin{equation}
	\label{eq:mutualJ}
	 J(A:B) = S(\rho_{B}) - S(B|A), 
\end{equation}
where the conditional entropy $S(B|A)$ is defined with respect to the measurement ${\cal M}$ performed on system $A$ with possible results labelled by $a_{k}$ as
\begin{equation}
	\label{eq:conditional1}
	 S(B|A) = \sum_{k} p_{k} S(\rho_{B}|a_{k}). 
\end{equation}
Here $p_{k} \equiv p(a_{k})$ is the probability of obtaining the measurement result $a_{k}$. Note that if $A$ and $B$ are assumed to the classical random variables with associated probability distributions then the classical analogues of the mutual informations $I(A:B)$ and $J(A:B)$ defined by replacing the vonNeumann entropies with the corresponding Shannon entropies in Eqs.~(\ref{eq:mutualI}) and (\ref{eq:mutualJ}) are identical as a consequence of Bayes' theorem. Subtracting $J$ from $I$ quantifies the correlations in the $AB$ system that are not revealed by the measurement ${\cal M}$ on $A$. Removing the ambiguity in $I-J$ stemming from the choice of measurement ${\cal M}$ by maximising over all possible local measurements, one defines discord as
\begin{equation}
	\label{eq:discord1}
	{\cal D} = I(A:B) - \max_{\cal M} J(A:B). 
\end{equation}
Note that if we replace the maximisation over all possible measurements on a subsystem with an operationally well defined measurement strategy, namely measurement in the eigen-basis of the local density operator, then we obtain another measure of nonClassical correlations, namely the measurement induced disturbance~\cite{luo08a}. In fact, even quantum discord is often defined in an operationally simpler way by limiting the maximisation in Eq.~(\ref{eq:discord1}) to one over all possible sets of projective measurements. In this paper we restrict ourselves to discord defined with respect to projective measurements and we explore one more dimension of the problem of quantifying nonClassical correlations by factoring in the disturbance to the measured system using the framework of weak quantum measurements. 

\section{Weak measurements \label{sec:weakmeasurements}}

The notion of post-measurement states in quantum mechanics is necessitated by the unavoidable disturbance to states due to measurements. Rather than leaving the post-measurement state to be defined by the details of the measurement that is performed, in the paradigm of weak measurements, from the outset, one defines a post measurement state $|\psi_{f} \rangle$ relative to the pre-measurement state $|\psi_{i} \rangle$ of a quantum system of interest. Within this two state-vector formalism of Aharnov, Albert and Vaidmann~\cite{Aharonov:PhysRevLett:1988}, the {\em weak value} of an observable ${\cal O}$ of the system is defined as
\begin {equation}
	\label{eq:weakvalue}
	\langle O\rangle_{w}=\frac{\left\langle \psi_{f}\left|{\cal O}\right|\psi_{i}\right\rangle}{\left\langle \psi_{f} |  \psi_{i}\right\rangle}.
\end{equation}
The weak value is peculiar because it can be larger than the eigenvalues of the observable. The expectation value of the weak operator can even lie outside the range of eigenvalues of the observable\cite{Dressel:2012dr,Dressel:2014ks,Dressel:PhysRevLett:2010}. 

Implementation of both projective and weak measurements can be framed in the language of pointer states~\cite{Zurek:1981cp,Zurek:2003fm,joos2013decoherence} by introducing a measuring device or {\em pointer} defined by a pair of canonically conjugate observables $Q$ and $P$. To measure the observable $A$, the system is coupled to the pointer through an interaction 
\begin{equation}
	H = g(t) {\cal O} \otimes P,
\end{equation}
in units where $\hbar = 1$. The function $g(t)$ indicates an interaction that occurs within a short span of time with $g(t)$ vanishing outside that interval. Since the magnitude of the disturbance produced by this interaction on the system is of interest to us, we can let the coupling act instantaneously and let 
\[ g(t) = g \delta (t-t_{0}). \]
Coupling the system and pointer together puts the two in the joint state $\exp(-ig {\cal O}\otimes P) |\psi_{i} \rangle |\Phi \rangle$, where $|\Phi  \rangle$ is the initial state of the pointer. Expanding the state of the system state in the eigenbasis $\{|a_{k} \rangle \}$ of ${\cal O}$ we find that the joint state of the system and the pointer after the interaction is
\[  \sum_{k} \psi_{ij} |a_{k} \rangle |\Phi (Q - ga_{k})\rangle, \]
where
\[ {\cal O} = \sum_{k} a_{k}|a_{k} \rangle \langle a_{k}| = \sum_{k} a_{k} \Pi_{k}.  \]
If the states $|\Phi(Q - ga_{k}) \rangle$ for different values of $a_{k}$ are such that they do not have significant overlap with each other then the measurement implemented by localising the pointer at a particular $Q$ will be projective on the system. On the other hand if $|\Phi(Q - ga_{k}) \rangle$ have significant overlap with each other, post-selection of the system state as $|\psi_{f} \rangle$ implements a weak measurement that ascribes the weak value $O_{w}$ given in Eq.~(\ref{eq:weakvalue}) to the observable ${\mathcal O}$. 

Let the $Q$-space representation of the initial state of the pointer be a minimum uncertainty gaussian wave packet around $Q=0$ with $\Delta Q = \Delta$. In the conjugate $P$-space, the uncertainty is $\Delta P = 1/2\Delta$. The significant overlap condition on $|\Phi(Q - ga_{k}) \rangle$ for implementing a weak measurement translates to the condition
\begin{equation}
	\label{eq:weakcondition1}
	g \Delta P = \frac{g}{2 \Delta} \ll \frac{|\langle \psi_{f} | \psi_{i} \rangle |}{|\langle \psi_{f}| {\cal O}^{n} | \psi_{i} \rangle|^{1/n}}, \quad {\rm for} \quad n=1,2, \ldots. 
\end{equation}

The weak value in Eq.~(\ref{eq:weakvalue}) is defined assuming pure initial states of the system followed by post selection through projective measurements. We require a generalisation of the weak value, weak measurement condition and the expression for the pointer shifts in the case where the initial state of the system is mixed subject to arbitrary post selection. Such a generalisation is possible~\cite{Wu:PhysicalReviewA:2011,Wu:PhysicsLettersA:2009,Wiseman:2002jt,Dressel:2012dr}, where the weak value is defined as
\begin{equation}
	\label{eq:weakvalue2}
	\langle O \rangle_{w} = \frac{{\rm tr} (P_{f} {\cal O} \rho_{i})}{{\rm tr}(P_{f} \rho_{i})},
\end{equation}
where $\rho_{i}$ is the initial state of the system and $P_{f}$ is a positive operator. Assuming that $\rho_{f}$ is not trace orthogonal to $\rho_{i}$, the weak-measurement condition in Eq.~(\ref{eq:weakcondition1}) is generalised to
\[ g\Delta P = \frac{g}{2\Delta} \ll 1. \]
The case where ${\rm tr}(\rho_{f} \rho_{i}) = 0$ has to be treated separately~\cite{Wu:PhysicalReviewA:2011} and it is not reprised here since it does not apply to our results. 

\section{Weak quantum discord \label{sec:weakdiscord}}

On a bipartite system in the state $\rho_{AB}$, imagine that weak measurements are performed on subsystem $A$ with the aim of collecting such information about $A$ as that would reveal as much of the classical correlations between $A$ and $B$. Minimising the disturbance on $A$ as well as on the overall system $AB$ due to the measurement being the motivation for performing weak measurements as opposed to strong measurements. Let $\rho_{AB}$ be the initial state of the system and as described in Sec.~\ref{sec:discord}, let $\{\Pi_{k}^{A} \}$, $k=1,\ldots,d_{A}$ be a set of orthogonal projectors that maximise $J(A:B)$ and thereby give the value of discord ${\cal D}(A \rightarrow B)$ in Eq.~(\ref{eq:discord1}). The post-selection that is part of the weak measurement is chosen to be on to a POVM element, 
\begin{equation}
	\label{eq:rhofinal1}
	P_{f} = (1 -\alpha) \rho_{AB} + \alpha \sum_{k=1}^{d_{A}}  \Pi_{k}^{A} \otimes \openone_{d_{B}}  = (1 -\alpha) \rho_{AB} + \alpha \openone_{d_{AB}}, 
\end{equation}
where $d_{A}$, $d_{B}$ and $d_{AB}$ are the dimensions of the measured subsystem $A$, subsystem $B$ and the whole system $AB$ respectively. Note that $P_{f}$ along with $P_{f}' = (1-\alpha)(\openone_{d_{AB}} - \rho_{AB})$ are two positive operators that together form a POVM that can be implemented on $A$. 

The projector $P_{f}$ is chosen so that when $\alpha$ is zero, then the weak measurement does not disturb the state of the system $AB$ at all while $\alpha =1$ corresponds to performing a complete set of projective measurements on subsystem $A$ on to a basis that maximises $J(A:B)$. Other values of the free parameter $\alpha$ interpolates smoothly between these two operations. This choice of post selection allows us to study the trade off between the disturbance to the state of $AB$ due to the measurement on $A$ and the degree to which the classical correlations between $A$ and $B$, as quantified by $J(A:B)$ are revealed by the results of weak measurement on $A$.

The projective measurements on to the set $\{ \Pi_{k}^{A} \}$ yield the probabilities 
\begin{equation}
	\label{eq:probs}
	p_{k} = {\rm tr}[(\Pi_{k}^{A} \otimes \openone_{d_{B}}  )\rho_{AB}],
\end{equation}
 for the various measurement outcomes, labeled $a_{k}$, for subsystem $A$. These probabilities, in turn, appear in the conditional entropy $S(B|A)$ in Eq.~(\ref{eq:conditional1}). We extend the definition of quantum discord so that it applies to the case where weak measurements are done on subsystem $A$ as opposed to projective measurements by estimating these probabilities using weak measurements. To make the connection, we choose the operator ${\mathcal O}$ appearing in Eq.~(\ref{eq:weakvalue2}) as 
\[ {\mathcal O} = \sum_{k=1}^{d_{A}} a_{k} \Pi_{k}^{A} \otimes \openone_{d_{B}} . \]
The eigenvalues $a_{k}$ are chosen so that the set of equations
\begin{equation}
	\langle {\mathcal O}^{n} \rangle = \sum_{k=1}^{d_{A}} a_{k}^{n} p_{k}, \qquad n =0,\ldots, d_{A}-1
\end{equation}
is invertible using Gaussian elimination or a similar standard algorithm. Here 
\[ \langle {\mathcal O}^{n} \rangle \equiv {\rm tr}({\mathcal O}^{n} \rho_{AB}), \]
is the expectation value of the $n^{\rm th}$ power of ${\mathcal O}$ with respect to the initial state of the system. 

We can use as an estimate of the probabilities for various measurement outcomes $a_{k}$ on subsystem $A$ the `weak' probabilities $p_{k}^{w}$ that are obtained by inverting the set of equations 
\begin{equation}
	\langle {\mathcal O}^{n} \rangle_{w} = \sum_{k=1}^{d_{A}} a_{k}^{n} p_{k}^{w}, \qquad n =0,\ldots, d_{A}-1
\end{equation}
where $\langle {\mathcal O}^{n}\rangle_{w}$ is the weak value of ${\mathcal O}^{n}$ as defined in Eq.~(\ref{eq:weakvalue2}) using post selection on to the POVM element in Eq. (\ref{eq:rhofinal1}). Note that $p_{k}^{w} = p_{k}$ when $\alpha = 1$. We can now define the {\em weak quantum discord} of the bipartite state $\rho_{AB}$ as
\begin{equation}
	\label{eq:weakdiscord}
	{\mathcal D}_{w} = S(\rho_{A}) - S(\rho_{AB}) + \sum_{j} p_{k}^{w} S(\rho_{B}|a_{k}).  	
\end{equation}
The probability $p_{k}$ for measurement outcome $a_{k}$ on $A$ is replaced by the `weak' version $p_{k}^{w}$. The conditioned entropy $S(\rho_{B}|a_{k})$ remains unchanged since the conclusion from the weak measurement remains that measurement outcome $a_{k}$ on $A$ indeed happens with probability $p_{k}$. Note that when $\alpha = 1$, then $P_{f} = \openone_{d_{AB}}$ and $\langle {\mathcal O}^{n} \rangle_{w} = \langle {\mathcal O}^{n} \rangle$ as can be seen from Eq.~(\ref{eq:weakvalue2}) and weak quantum discord reduces to normal discord as required. 

\subsection{Measurements on qubits \label{qubitmeasure}}

If the measured system $A$ is restricted to being a single qubit - as is the case in all the examples we will be considering in detail below - only the weak value of the operator
\begin{equation}
	\label{eq:qubitop1}
	{\mathcal O} =  (\Pi^{A}_{+} - \Pi^{A}_{-}) \otimes \openone_{B} , 
\end{equation}
suffices to estimate the corresponding probabilities $p_{+}^{w}$ and $p_{-}^{w}$. Here $\Pi^{A}_{\pm}$, as mentioned earlier, are orthogonal projectors that maximise the quantum discord in Eq.~(\ref{eq:discord1}). Using $p_{+}^{w}+p_{-}^{w}=1$ and $p_{+}^{w} - p_{-}^{w} = \langle {\mathcal O} \rangle_{w}$, we have
\[ p_{\pm}^{w} = \frac{1 \pm \langle {\mathcal O} \rangle_{w}}{2}. \]
 
Since $\langle {\mathcal O} \rangle_{w}$ alone determines the weak discord when the measured system is qubit, it is worth looking at the weak value in a bit more detail. We have
 \[ {\rm tr}(P_{f} {\mathcal O} \rho_{AB} ) = (1-\alpha){\rm tr}({\mathcal O} \rho_{AB}^{2}) + \alpha \langle {\mathcal O} \rangle, \]
 and
 \[ {\rm tr}(P_{f} \rho_{AB}) = (1-\alpha) {\rm tr}(\rho_{AB}^{2}) + \alpha, \]
so that
\begin{equation}
	\label{eq:qubitweak} 
	\langle {\mathcal O} \rangle_{w} = \frac{ (1-\alpha){\rm tr}({\mathcal O} \rho_{AB}^{2}) + \alpha \langle {\mathcal O} \rangle}{(1-\alpha) {\rm tr}(\rho_{AB}^{2}) + \alpha}. 
\end{equation}
From Eq.~(\ref{eq:qubitweak}) we find that for all states for which 
\begin{equation}
	\label{eq:condition1}
	 {\rm tr} ({\mathcal  O} \rho_{AB}^{2}) = \langle {\mathcal O} \rangle {\rm tr}(\rho_{AB}^{2}), 
\end{equation}
the weak value of ${\mathcal O}$ and its expectation value $\langle {\mathcal O} \rangle$ coincide when the measured system is a qubit. In particular this is true for all pure states of the bipartite system $AB$ since $\rho_{AB}^{2} = \rho_{AB}$. In general if $\rho_{AB}^{2} = k \rho_{AB}$, where $k$ is a constant, then also condition (\ref{eq:condition1}) is satisfied. 

\subsection{Bell Diagonal States \label{sec:belldiagonal}}

Bell diagonal states are two qubit states of the form~\cite{Horodecki:1996fc,Horodecki:2009gb}
\begin{equation}
	\label{eq:belldiagonal}
{\rho}_{AB} = \frac{1}{4}\bigg(\openone \otimes \openone + \sum^{3}_{j=1}c_{j}\sigma^{A}_{j} \otimes \sigma^{B}_{j}\bigg),
\end{equation}
where $\sigma_{j} $'s are the Pauli matrices and $\left|c_{1}\right|+\left|c_{2}\right|+\left|c_{3}\right| \leq  1$ so that $\rho_{AB}$ is positive semi-definite. These states have the four Bell states, 
\[ |\beta_{ab} \rangle = \frac{1}{\sqrt{2}} (|0,b\rangle + (-1)^{a}|1,1\oplus b \rangle), \qquad a,b=0,1,\]
as eigenstates and maximally mixed marginal density operators so that $S(\rho_{A}) = S(\rho_{B}) = 1$. The eigenvalues of the Bell diagonal states are
\begin{equation}
	\label{eq:belleigs}
	\lambda_{ab} = \frac{1}{4}[1 + (-1)^{a}c_{1} - (-1)^{a+b} c_{2} + (-1)^{b}c_{3}].
\end{equation}
The quantum mutual information shared between the two qubits is
\begin{equation}
	\label{eq:bellmutualI}
	I(A:B) = 2 +\sum_{a,b}\lambda_{ab} \log \lambda_{ab}=\sum_{a,b}\lambda_{ab} \log (4 \lambda_{ab}).
\end{equation}
Quantum discord in Bell-diagonal states can be computed~\cite{Lang:PhysRevLett:2010} using
\begin{equation}
	\label{eq:bellmutualJ}
	\max_{\cal M} J(A:B) = \frac{1}{2}[(1+c_{s})\log(1+c_{s}) + (1-c_{s})\log (1-c_{s})], 
\end{equation}
where $c_{s} \equiv \max |c_{j}|$. The pair of orthogonal projectors that maximise $J(A:B)$ are 
\[ \Pi_{\pm}^{A} = \frac{1}{2}(\openone \pm \sigma_{s}), \]
where the index $s$ indicates the direction in Bloch sphere corresponding to $\max |c_{j}|$. So we have 
\[ {\mathcal O} = \sigma_{s}^{A} \otimes \openone_{B}. \]
Using the fact that the Pauli matrices are traceless, it is easy to see that
\[ \langle {\mathcal O} \rangle = {\rm tr}[(\sigma_{s}^{A} \otimes \openone_{B}) \rho_{AB} ] = 0. \]
We also have
\[ \rho_{AB}^{2} \!\!=\!\! \frac{1}{16} \bigg[ \big(1 + \sum_{j} c_{j}^{2} \big) \openone \otimes \openone + 2 \!\! \sum_{j} (c_{j} - |\epsilon_{jkl}| c_{k}c_{l}) \sigma_{j}^{A} \otimes \sigma_{j}^{B} \bigg], \]
from which it follows that
\[ {\rm tr}({\mathcal O} \rho_{AB}^{2}) = 0.\]
So we find that for Bell-diagonal states, 
\[ \langle {\mathcal O} \rangle_{w} = \langle {\mathcal O} \rangle = 0, \]
and 
\[ p_{\pm}^{w} = p_{\pm} = \frac{1}{2}. \]
The above results mean that the weak quantum discord is identical to the normal quantum discord for Bell-Diagonal states. 

The Werner state,
\[  {\rho}_{AB} = \frac{1}{4}\bigg(\openone \otimes \openone + c\sum^{3}_{j=1} \sigma^{A}_{j} \otimes \sigma^{B}_{j}\bigg), \]
is a special case of the Bell-Diagonal family of states with $c_{1} = c_{2} = c_{3} = c$. The isotropic nature of the state means that any pair or orthogonal projectors on to mutually orthogonal directions on the Bloch sphere will maximise $J_{AB}$ can any such pair can be used to define the discord and weak discord. As outlined above, in the case of Werner states also, irrespective of the value of $c$, the weak quantum discord is identical to the normal discord. 

\subsection{Randomly generated two qubit states}

The two examples in the previous section are ones for which the weak quantum discord and normal quantum discord coincide. While this seems to be true for several well studied families of bipartite states for which the discord is computed by measuring a one-qubit subsystem, for arbitrary states it is not so. We do not expect the weak discord to be either less than or greater than the regular discord since discord itself is neither convex or concave over the set of states. In defining the weak quantum discord we are replacing the conditional entropy $S(B|A)=\sum_{k} p_{k} S(B|a_{k})$ with $\sum_{k} p_{k}^{w} S(B|a_{k})$ where only the $p_{k}$'s have been replaced by $p_{k}^{w}$'s.  Whether the conditional entropy, and consequently the weak discord, increases or decreases relative to the normal discord as a result of this replacement depends on the values of $S(B|a_{k})$ for $k=1,\dots,d_{A}$.  

\begin{figure}[!htb]
	\resizebox{8.2cm}{5.5cm}{\includegraphics{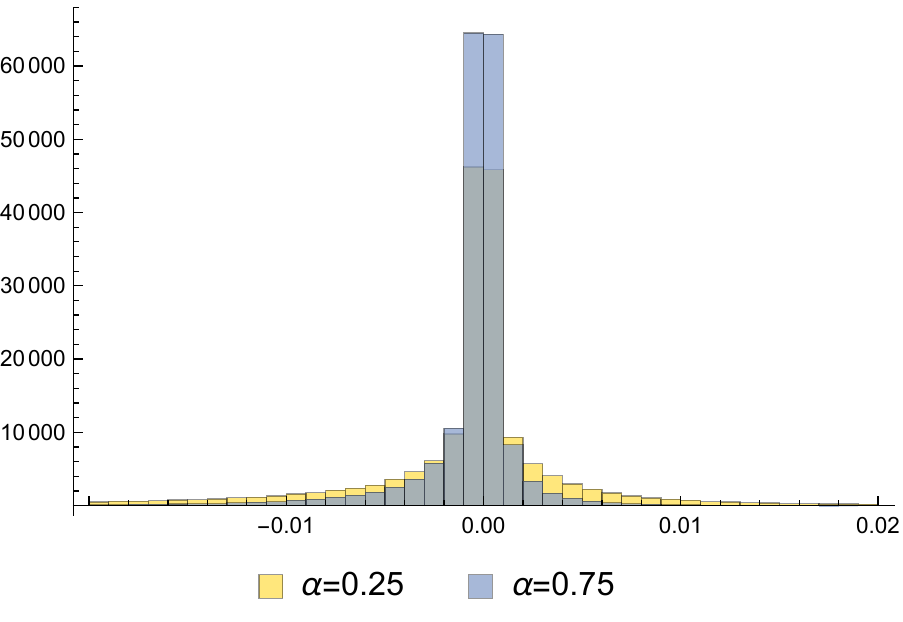}}
	\caption{Histograms showing the distribution of ${\cal D}^{w} - {\cal D}$ for 175,000 randomly generated two qubit mixed states. The histogram in the background (yellow) shows the distribution for $\alpha = 0.25$ while the one in the foreground (blue) is for $\alpha = 0.75$. We see that the weak discord is in general close to the value of the normal discord and distributed on either side of the value of normal discord with a very narrow spread. The spread decreases as expected when $\alpha$ increases since when $\alpha$ tends to one, the weak discord reduces to the normal quantum discord.  \label{fig1}}
\end{figure}

The distribution of the difference between the weak discord and normal discord computed for around 175,000 randomly generated two qubit states is plotted as a histogram in Fig.~\ref{fig1}. The two qubit states of ranks two, three or four (for rank one pure states, the weak discord and normal discord coincide), were generated by first creating a $4\times4$, diagonal matrix of unit trace with two, three or four randomly picked entries along the diagonal followed by the application of a random unitary transformation. The orthogonal projectors that give the quantum discord for each randomly generated state was obtained by numerically minimising over the set of all pairs of such projectors which are parametrised by two angles. Using these projectors, the weak quantum discord was computed following the steps described earlier. 

\begin{figure}[!htb]
	\resizebox{8.2cm}{5.5cm}{\includegraphics{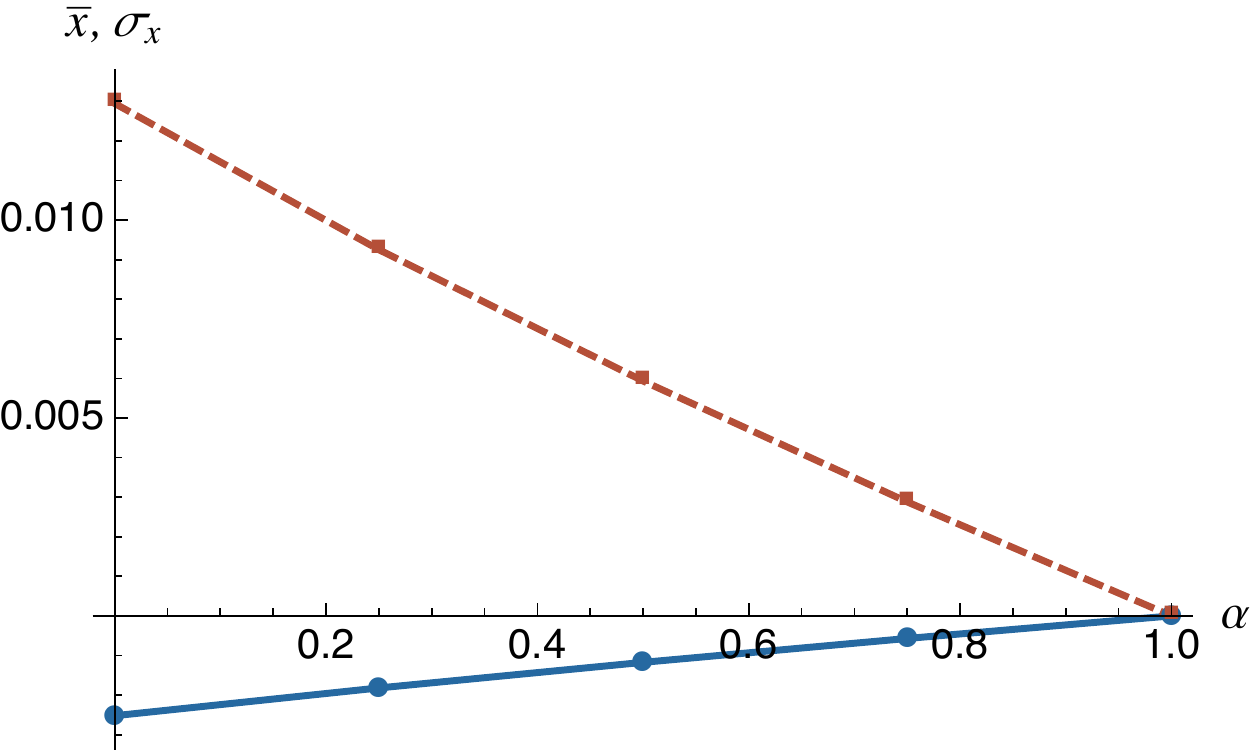}}
	\caption{ The solid (blue) line shows the dependence of the mean of ${\mathcal D}_{w} - {\mathcal D}$ averaged over 175,000 instances of randomly generated states on the parameter $\alpha$ appearing in the definition of the post selection POVM, $P_{f}$. The dashed (red) line shows the dependence on $\alpha$ of the standard deviation of the distribution of ${\mathcal D}_{w} - {\mathcal D}$. \label{fig2}}
\end{figure}

Fig.~\ref{fig1} contains two histograms corresponding to the distribution of the difference ${\cal D}^{w} - {\cal D}$ for two different values of the parameter $\alpha$ that appears in the post-selection POVM element $P_{f}$. We have chosen $P_{f}$ such that when $\alpha \rightarrow 1$ then $\langle {\mathcal O} \rangle_{w} \rightarrow \langle {\mathcal O} \rangle$ and ${\mathcal D}_{w} \rightarrow {\mathcal D}$. Numerically we find that as $\alpha$ increases, the values of weak discord are clustered more narrowly around the value of the normal discord. The mean of the difference ${\cal D}^{w} - {\cal D}$ averaged over all the 175,000 randomly generated states as well as the the standard deviation of the distribution of the difference is plotted in Fig.~\ref{fig2}. Apart from noting that the values of ${\mathcal D}_{w}$ are more sharply clustered around ${\mathcal D}$ for larger values of $\alpha$ we also note that on an average the weak quantum discord is lesser than the normal quantum discord by a very small amount. This means that using weak measurements instead of projective measurements on subsystem $A$ over-estimates the amount of classical correlations between $A$ and $B$ by a very small amount (less than 2 percent for $\alpha=0.25$) on an average. We defer a detailed discussion on the closeness of weak quantum discord and normal discord to Sec.~\ref{sec:conclusion}.

\subsection{The DQC1 state \label{DQC1}}

The DQC1 model of quantum computation~\cite{Knill:PhysRevLett:1998}, uses a single qubit with finite purity coupled to a register of $n$ qubits in the fully mixed state $\openone_{d_{n}}/2^{n}$ to evaluate the normalised trace of a random unitary matrix acting on $n$ qubits exponentially more efficiently in comparison with the best known classical algorithm. DQC1 is also now recognised as belonging to a separate complexity class among computational problems.  The interest in DQC1 in the context of our discussion on weak quantum discord stems from it being a viable and well studied model of mixed state quantum computation. Additionally, it has been observed that while entanglement between the pure qubit and the register of mixed qubits is zero during all stages of the computation, quantum discord is generated between the two subsystems during the computation~\cite{Datta:PhysRevLett:2008}. 

Given the unitary matrix $U$ of dimension $2^{n}$ and an efficient implementation of the unitary as a sequence of quantum gates on the $n$-qubit register, application of the unitary controlled on the initial state $(|0\rangle + |1\rangle)/2$ of the pure qubit, puts the DQC1 system of $n+1$ qubits in the state
\begin{eqnarray}
	 \rho_{AB} & = &  \frac{1}{2^{n+1}} \big( |0\rangle \langle 0| \otimes \openone_{d_{n}} + |1\rangle \langle 1| \otimes \openone_{d_{n}}  \nonumber \\
	 && \qquad \qquad + |0\rangle \langle 1| \otimes U^{\dagger} + |1\rangle \langle 0| \otimes U \big).
	 \label{eq:dqc1state1}
\end{eqnarray}
It is easy to see that
\[ \rho_{AB}^{2} = \frac{1}{2^{n}} \rho_{AB}, \]
and so it follows that for the DQC1 state ${\mathcal D}_{w} = {\mathcal D}$. 

\subsection{A possible alternative}

An alternative definition of the weak quantum discord is worth considering for completeness of our discussion. Given the POVM element $P_{f}$ that corresponds to the post-selection applied in the weak measurement scheme, we can assign the following post measurement state to the bipartite system after the measurement, 
\begin{equation}
	\label{eq:postmeasure1}
	\rho_{AB}' = \frac{P_{f} \rho_{AB} P_{f}^{\dagger}}{ {\rm tr}(P_{f} \rho_{AB} P_{f}^{\dagger})}. 
\end{equation}
The post measurement state of sub-system $B$ is therefore
\[ \rho_{B}' = {\rm tr}_{A} (\rho_{AB}'). \]
We may then choose to define the mutual information between $B$ and $A$ conditioned on a weak measurement on $A$ as
\[J(A:B) = S(\rho_{B}) - S(\rho_{B}'),\]
and the weak discord as ${\mathcal D}^{w} = I(A:B) - J(A:B)$, which, in turn, is a quantity independent of the projective measurements that appear in the definition of normal quantum discord. However this possible way of defining the weak quantum discord lays emphasis on the disturbance to the system and does not factor in what, if any, is known about the sub-system upon doing the measurement. Since quantum discord is about correlations, both classical and nonClassical, between $A$ and $B$ that is revealed when one of the subsystems is measured, we choose the definition of weak discord given in Eq.~(\ref{eq:weakdiscord}). 

\section{Interpretations of discord and weak measurements  \label{sec:interpretations}}

Our observations on weak measurements in relation to nonClassical correlations in quantum states raise interesting questions on how we understand quantities like the quantum discord. As mentioned in the introduction, Ollivier and Zurek~\cite{Ollivier:PhysRevLett:2001} introduced discord as a means of characterising the disturbance on one subsystem of a bipartite state due to projective measurements on the other. Zero discord states are of the form 
\[ \rho_{AB} = \sum_{k} \rho^{B}_{k} \otimes \Pi_{k}^{A}, \]
and projective measurements on $A$ using $\Pi_{k}^{A}$ do not affect the state of sub-system $B$. If the objective of the projective measurement on one of the sub-systems is only to `know' the measurement statistics corresponding to the complete set of orthogonal projectors $\{\Pi_{k}^{A} \}$, then our analysis indicates that for a larger class of states than the set of zero discord states, we can obtain the very same statistics with arbitrarily small disturbance not only to sub-system $B$ but to the overall bipartite state as well. 

In the weak measurement scheme discussed, the post selection of the bipartite state using the positive operator $P_{f}$ does constitute a disturbance to both the measured sub-system as well as to the overall system. We can quantify the disturbance using the probability $p_{f}$ for obtaining a positive outcome to the post-selection, 
\[ p_{f} = {\rm tr}(P_{f} \rho_{AB} P_{f}^{\dagger}) . \] 
For all states $\rho_{AB}$ characterised by $\rho_{AB}^{2} = k\rho_{AB}$ it is easy to show that
\[ p_{f}  = \big[k(1-\alpha) + \alpha \big]^{2}, \]
The post measurement state after the application of the POVM for such states is 
\[ \rho_{AB}' = \frac{P_{f} \rho_{AB} P_{f}^{\dagger}}{ {\rm tr}(P_{f} \rho_{AB} P_{f}^{\dagger})} = \rho_{AB}. \]
This means that the disturbance on sub-system $B$ due to the weak measurement which estimates the outcome statistics, $p_{k}$, corresponding to $\Pi_{k}^{A}$ accurately as $p_{k}^{w}$ can be made arbitrarily small whenever $k$ is close to unity by choosing $\alpha$ to be very small.  Hence identifying the set of zero discord states with the set of states on which measurements on $A$ does not disturb sub-system $B$ is called into question provided the measurements are understood to be ones designed to reveal the statistics of various outcomes corresponding to a complete set of projective measurements on to $\{\Pi_{k}^{A} \}$.

In~\cite{Zurek:2003el}, Zurek discusses a thermodynamic approach to understanding quantum discord. A slightly modified version of discord is shown to be equal to the difference between the efficiency of quantum and classical Maxwell's demons in extracting work from correlated quantum systems. The quantum demon can do non-local measurements and therefore has access to all the correlations, quantum or otherwise, in a bipartite state. The classical demon can only do local measurements. Given a bipartite state $\rho_{AB}$, the classical demon measures the state of subsystem $A$ and uses the information obtained about $B$ to extract work from the subsystem $B$ by letting it expand through the available Hilbert space dimension $d_{B}$ while in contact with a reservoir at temperature $T$. Working in units of $k_{B_{2}}T$, where $k_{B_{2}}$ is the Boltzmann's constant scaled so as to use logarithms to base two rather than natural logs, the work extracted by the local classical demon from $B$ is obtained as 
\[ W^{+}_{c} = \log d_{B} - S(B|\{\Pi_{k}^{A} \}). \]
In the expression above the conditional entropy is computed as discussed earlier conditioned on the results obtained by the demon on measuring $A$. We may as well assume that the classical demon has already optimised the measurements to extract maximum possible work from $B$. Landauer's principle~\cite{Landauer:1961kg} highlights the need to reset the demon's memory to bring the system back to the initial state for repeating the process of extracting work from another copy of the state $\rho_{AB}$. The demon's memory contains the probabilities $p_{k}$ of various measurement outcomes on $A$ and the classical memory required to hold the $p_{k}$ is of the order of the dimension $d_{A}$ of $A$. The classical demon can however compress this data depending on the nature of the probabilities $p_{k}$ and the net cost, in terms of work, for erasure of the demon's memory is
\[ W^{-}_{c} = \log d_{A} -  S(A). \]
The total work extracted by the classical demon is therefore
\[ W_{c} = \log d_{AB} - [S(A) + S(B|\Pi_{k}^{A})]. \]
 On the other hand, the quantum demon who can make observations in the full Hilbert space of $AB$ can extract work 
 \[ W_{q} = \log d_{AB} - S(A,B). \]
 We see that indeed quantum discord is equal to $W_{q} - W_{c}$. 
 
 This interpretation of the quantum discord goes across relatively unscathed when we expand its scope to the case of weak quantum discord, primarily because it emphasises what can be done by the classical demon with $B$ given the knowledge of $A$ and does not depend crucially on how the knowledge about $A$ is acquired. We have seen that the weak measurements on $A$ provide the same information as projective measurements in many cases and in others, it provides almost the same information by giving good estimates $p_{k}^{w}$ of $p_{k}$. The work that the classical demon can extract from $B$ is modified slightly, if at all, in replacing $p_{k}$ with $p_{k}^{w}$ when computing the conditional entropy $S(B|\Pi_{k}^{A})$. There may be a slight change in the cost of erasure of the demon's memory also since the algorithmic complexity~\cite{li2013introduction} of $p_{k}$ may be different from that of $p_{k}^{w}$. 

A pair of operational interpretations of quantum discord based on information theoretic protocols were given by Cavalcanti et al.~\cite{Cavalcanti:2011dy} and Madhok et al.~\cite{Madhok:2011bh}. Since the connection between the two interpretations is discussed in the respective references themselves, we will restrict our discussion to the operational interpretation given in~\cite{Madhok:2011bh}. The protocol used in~\cite{Madhok:2011bh} is quantum state merging. Observers $A$ and $B$ hold $n$ copies of the tripartite state $|\psi_{ABC} \rangle$ which is the purification of the bipartite state $\rho_{AB}$ that is shared across $A$ and $B$. Quantum state merging protocol quantifies the minimum amount of quantum information that $B$ should transfer to $A$ so that $A$ who is assumed to hold $C$ as well, can re-create the state $\rho_{AA'}$ (or equivalently $|\psi_{AA'C} \rangle$) which is arbitrarily close to $\rho_{AB}$ entirely at her end. The minimum amount of quantum information to be transferred is quantified by the quantum conditional entropy $S(B|A) = S(A,B)-S(A)$ (without measurements on $A$). The interpretation in~\cite{Madhok:2011bh} identifies quantum discord as the markup in the quantum communication needed from $B$ to $A$ to do state merging in case $A$ chooses to measure her state before state merging. 

For the operational interpretation based on quantum state merging to hold when weak measurements are involved, one has to further clarify what is meant by $A$ `measuring' the state of her sub-system. If the measurement is done in the sense that the probabilities of various outcomes are estimated, then this operational interpretation does not apply in the case where weak measurements are employed since a concurrent collapse of the state of $A$ to one of the basis states is also required for the interpretation to apply. In establishing the operational interpretation, sub-system $C$ essentially takes on the role of the pointer. In particular, $|\psi_{ABC} \rangle$ is viewed in~\cite{Madhok:2011bh} as a purification of $\rho_{AB}$ created from a product state of $AB$ and $C$ through a unitary interaction between $A$ and $C$. This means that during the state merging protocol, $C$ behaves like the pointer as applicable to strong (projective) measurements with mutually orthogonal states of $C$ being associated with possible measurement outcomes on $A$. Discarding $C$ as described in~\cite{Madhok:2011bh} implements the projective measurement on $A$ for the ensemble of $n$ copies of $|\psi_{ABC} \rangle$. This scenario is not applicable when weak measurements on $A$ are used to estimate measurement statistics. Hence this operational interpretation does not extend naturally to the case of weak quantum discord. 

\section{Conclusion \label{sec:conclusion}}

We have introduced an extension of quantum discord to the case where weak measurements are employed to estimate the measurement statistics of one of the sub-systems and defined the weak quantum discord. We find that for several families of states, the weak discord coincides with the normal discord in addition to the weak discord reducing to the normal discord in the limit where the weak measurements turn into projective ones. Even when the two do not coincide we find that for a large number of two qubit states for which both are evaluated numerically, the weak discord values are very close to the normal discord values. 

Interpretations of quantum discord, operational or otherwise, do not extend over to weak discord whenever the interpretation lays emphasis on the measurement and its nature on one of the sub-systems. The numerical closeness of weak discord to normal discord therefore strongly suggests that interpretations that focus on what more (or less) can be done by knowing the statistics of measurements on one of the subsystems are desirable for normal quantum discord also. An independent operational interpretation of weak quantum discord remains to be developed. 

\acknowledgments

Lekshmi S. acknowledges the support of the University Grants Commission, Government of India, through the Faculty Development Program (No. KLKE001TF11). Anil Shaji acknowledges the support of the Department of Science and Technology, Government of India, through the Ramanujan Fellowship program (No. SR/S2/RJN-01/2009)

\bibliography{WeakDiscord}

\begin{thebibliography}{27}%
\makeatletter
\providecommand \@ifxundefined [1]{%
 \@ifx{#1\undefined}
}%
\providecommand \@ifnum [1]{%
 \ifnum #1\expandafter \@firstoftwo
 \else \expandafter \@secondoftwo
 \fi
}%
\providecommand \@ifx [1]{%
 \ifx #1\expandafter \@firstoftwo
 \else \expandafter \@secondoftwo
 \fi
}%
\providecommand \natexlab [1]{#1}%
\providecommand \enquote  [1]{``#1''}%
\providecommand \bibnamefont  [1]{#1}%
\providecommand \bibfnamefont [1]{#1}%
\providecommand \citenamefont [1]{#1}%
\providecommand \href@noop [0]{\@secondoftwo}%
\providecommand \href [0]{\begingroup \@sanitize@url \@href}%
\providecommand \@href[1]{\@@startlink{#1}\@@href}%
\providecommand \@@href[1]{\endgroup#1\@@endlink}%
\providecommand \@sanitize@url [0]{\catcode `\\12\catcode `\$12\catcode
  `\&12\catcode `\#12\catcode `\^12\catcode `\_12\catcode `\%12\relax}%
\providecommand \@@startlink[1]{}%
\providecommand \@@endlink[0]{}%
\providecommand \url  [0]{\begingroup\@sanitize@url \@url }%
\providecommand \@url [1]{\endgroup\@href {#1}{\urlprefix }}%
\providecommand \urlprefix  [0]{URL }%
\providecommand \Eprint [0]{\href }%
\providecommand \doibase [0]{http://dx.doi.org/}%
\providecommand \selectlanguage [0]{\@gobble}%
\providecommand \bibinfo  [0]{\@secondoftwo}%
\providecommand \bibfield  [0]{\@secondoftwo}%
\providecommand \translation [1]{[#1]}%
\providecommand \BibitemOpen [0]{}%
\providecommand \bibitemStop [0]{}%
\providecommand \bibitemNoStop [0]{.\EOS\space}%
\providecommand \EOS [0]{\spacefactor3000\relax}%
\providecommand \BibitemShut  [1]{\csname bibitem#1\endcsname}%
\let\auto@bib@innerbib\@empty
\bibitem [{\citenamefont {Ollivier}\ and\ \citenamefont
  {Zurek}(2001)}]{Ollivier:PhysRevLett:2001}%
  \BibitemOpen
  \bibfield  {author} {\bibinfo {author} {\bibfnamefont {H.}~\bibnamefont
  {Ollivier}}\ and\ \bibinfo {author} {\bibfnamefont {W.~H.}\ \bibnamefont
  {Zurek}},\ }\href@noop {} {\bibfield  {journal} {\bibinfo  {journal} {Phys.
  Rev. Lett.}\ }\textbf {\bibinfo {volume} {88}},\ \bibinfo {pages} {017901}
  (\bibinfo {year} {2001})}\BibitemShut {NoStop}%
\bibitem [{\citenamefont {Henderson}\ and\ \citenamefont
  {Vedral}(2001)}]{Henderson:JournalOfPhysicsAMathematicalAndGeneral:2001}%
  \BibitemOpen
  \bibfield  {author} {\bibinfo {author} {\bibfnamefont {L.}~\bibnamefont
  {Henderson}}\ and\ \bibinfo {author} {\bibfnamefont {V.}~\bibnamefont
  {Vedral}},\ }\href {\doibase 10.1088/0305-4470/34/35/315} {\bibfield
  {journal} {\bibinfo  {journal} {J. Phys. A:Math Gen}\ }\textbf {\bibinfo
  {volume} {34}},\ \bibinfo {pages} {6899} (\bibinfo {year}
  {2001})}\BibitemShut {NoStop}%
\bibitem [{\citenamefont {Modi}\ \emph {et~al.}(2012)\citenamefont {Modi},
  \citenamefont {Brodutch}, \citenamefont {Cable}, \citenamefont {Paterek},\
  and\ \citenamefont {Vedral}}]{Modi:ReviewsOfModernPhysics:2012}%
  \BibitemOpen
  \bibfield  {author} {\bibinfo {author} {\bibfnamefont {K.}~\bibnamefont
  {Modi}}, \bibinfo {author} {\bibfnamefont {A.}~\bibnamefont {Brodutch}},
  \bibinfo {author} {\bibfnamefont {H.}~\bibnamefont {Cable}}, \bibinfo
  {author} {\bibfnamefont {T.}~\bibnamefont {Paterek}}, \ and\ \bibinfo
  {author} {\bibfnamefont {V.}~\bibnamefont {Vedral}},\ }\href {\doibase
  10.1103/RevModPhys.84.1655} {\bibfield  {journal} {\bibinfo  {journal}
  {Reviews of Modern Physics}\ }\textbf {\bibinfo {volume} {84}},\ \bibinfo
  {pages} {1655} (\bibinfo {year} {2012})}\BibitemShut {NoStop}%
\bibitem [{\citenamefont {Datta}\ \emph {et~al.}(2008)\citenamefont {Datta},
  \citenamefont {Shaji},\ and\ \citenamefont {Caves}}]{Datta:PhysRevLett:2008}%
  \BibitemOpen
  \bibfield  {author} {\bibinfo {author} {\bibfnamefont {A.}~\bibnamefont
  {Datta}}, \bibinfo {author} {\bibfnamefont {A.}~\bibnamefont {Shaji}}, \ and\
  \bibinfo {author} {\bibfnamefont {C.~M.}\ \bibnamefont {Caves}},\ }\href@noop
  {} {\bibfield  {journal} {\bibinfo  {journal} {Phys. Rev. Lett.}\ }\textbf
  {\bibinfo {volume} {100}},\ \bibinfo {pages} {050502} (\bibinfo {year}
  {2008})}\BibitemShut {NoStop}%
\bibitem [{\citenamefont {Lang}\ \emph {et~al.}(2011)\citenamefont {Lang},
  \citenamefont {Caves},\ and\ \citenamefont
  {Shaji}}]{LANG:IntJQuanumInform:2011}%
  \BibitemOpen
  \bibfield  {author} {\bibinfo {author} {\bibfnamefont {M.~D.}\ \bibnamefont
  {Lang}}, \bibinfo {author} {\bibfnamefont {C.~M.}\ \bibnamefont {Caves}}, \
  and\ \bibinfo {author} {\bibfnamefont {A.}~\bibnamefont {Shaji}},\ }\href
  {\doibase 10.1142/s021974991100826x} {\bibfield  {journal} {\bibinfo
  {journal} {Int. J. Quanum Inform.}\ }\textbf {\bibinfo {volume} {09}},\
  \bibinfo {pages} {1553} (\bibinfo {year} {2011})}\BibitemShut {NoStop}%
\bibitem [{\citenamefont {Aharonov}\ \emph {et~al.}(1988)\citenamefont
  {Aharonov}, \citenamefont {Albert},\ and\ \citenamefont
  {Vaidman}}]{Aharonov:PhysRevLett:1988}%
  \BibitemOpen
  \bibfield  {author} {\bibinfo {author} {\bibfnamefont {Y.}~\bibnamefont
  {Aharonov}}, \bibinfo {author} {\bibfnamefont {D.~Z.}\ \bibnamefont
  {Albert}}, \ and\ \bibinfo {author} {\bibfnamefont {L.}~\bibnamefont
  {Vaidman}},\ }\href {\doibase 10.1103/PhysRevLett.60.1351} {\bibfield
  {journal} {\bibinfo  {journal} {Phys. Rev. Lett.}\ }\textbf {\bibinfo
  {volume} {60}},\ \bibinfo {pages} {1351} (\bibinfo {year}
  {1988})}\BibitemShut {NoStop}%
\bibitem [{\citenamefont {Duck}\ \emph {et~al.}(1989)\citenamefont {Duck},
  \citenamefont {Stevenson},\ and\ \citenamefont
  {Sudarshan}}]{Duck:PhysicalReviewD:1989}%
  \BibitemOpen
  \bibfield  {author} {\bibinfo {author} {\bibfnamefont {I.~M.}\ \bibnamefont
  {Duck}}, \bibinfo {author} {\bibfnamefont {P.~M.}\ \bibnamefont {Stevenson}},
  \ and\ \bibinfo {author} {\bibfnamefont {E.~C.~G.}\ \bibnamefont
  {Sudarshan}},\ }\href {\doibase 10.1103/PhysRevD.40.2112} {\bibfield
  {journal} {\bibinfo  {journal} {Phys. Rev. D}\ }\textbf {\bibinfo {volume}
  {40}},\ \bibinfo {pages} {2112} (\bibinfo {year} {1989})}\BibitemShut
  {NoStop}%
\bibitem [{\citenamefont {Singh}\ and\ \citenamefont
  {Pati}(2014)}]{singh2014quantum}%
  \BibitemOpen
  \bibfield  {author} {\bibinfo {author} {\bibfnamefont {U.}~\bibnamefont
  {Singh}}\ and\ \bibinfo {author} {\bibfnamefont {A.~K.}\ \bibnamefont
  {Pati}},\ }\href@noop {} {\bibfield  {journal} {\bibinfo  {journal} {Annals
  of Physics}\ }\textbf {\bibinfo {volume} {343}},\ \bibinfo {pages} {141}
  (\bibinfo {year} {2014})}\BibitemShut {NoStop}%
\bibitem [{\citenamefont {Luo}(2008)}]{luo08a}%
  \BibitemOpen
  \bibfield  {author} {\bibinfo {author} {\bibfnamefont {S.}~\bibnamefont
  {Luo}},\ }\href@noop {} {\bibfield  {journal} {\bibinfo  {journal} {Physical
  Review~A}\ }\textbf {\bibinfo {volume} {77}},\ \bibinfo {pages} {022301}
  (\bibinfo {year} {2008})}\BibitemShut {NoStop}%
\bibitem [{\citenamefont {Dressel}\ and\ \citenamefont
  {Jordan}(2012)}]{Dressel:2012dr}%
  \BibitemOpen
  \bibfield  {author} {\bibinfo {author} {\bibfnamefont {J.}~\bibnamefont
  {Dressel}}\ and\ \bibinfo {author} {\bibfnamefont {A.~N.}\ \bibnamefont
  {Jordan}},\ }\href@noop {} {\bibfield  {journal} {\bibinfo  {journal}
  {Physical Review A}\ }\textbf {\bibinfo {volume} {85}},\ \bibinfo {pages}
  {012107} (\bibinfo {year} {2012})}\BibitemShut {NoStop}%
\bibitem [{\citenamefont {Dressel}\ \emph {et~al.}(2014)\citenamefont
  {Dressel}, \citenamefont {Malik}, \citenamefont {Miatto}, \citenamefont
  {Jordan},\ and\ \citenamefont {Boyd}}]{Dressel:2014ks}%
  \BibitemOpen
  \bibfield  {author} {\bibinfo {author} {\bibfnamefont {J.}~\bibnamefont
  {Dressel}}, \bibinfo {author} {\bibfnamefont {M.}~\bibnamefont {Malik}},
  \bibinfo {author} {\bibfnamefont {F.~M.}\ \bibnamefont {Miatto}}, \bibinfo
  {author} {\bibfnamefont {A.~N.}\ \bibnamefont {Jordan}}, \ and\ \bibinfo
  {author} {\bibfnamefont {R.~W.}\ \bibnamefont {Boyd}},\ }\href@noop {}
  {\bibfield  {journal} {\bibinfo  {journal} {Reviews of Modern Physics}\
  }\textbf {\bibinfo {volume} {86}},\ \bibinfo {pages} {307} (\bibinfo {year}
  {2014})}\BibitemShut {NoStop}%
\bibitem [{\citenamefont {Dressel}\ \emph {et~al.}(2010)\citenamefont
  {Dressel}, \citenamefont {Agarwal},\ and\ \citenamefont
  {Jordan}}]{Dressel:PhysRevLett:2010}%
  \BibitemOpen
  \bibfield  {author} {\bibinfo {author} {\bibfnamefont {J.}~\bibnamefont
  {Dressel}}, \bibinfo {author} {\bibfnamefont {S.}~\bibnamefont {Agarwal}}, \
  and\ \bibinfo {author} {\bibfnamefont {A.~N.}\ \bibnamefont {Jordan}},\
  }\href@noop {} {\bibfield  {journal} {\bibinfo  {journal} {Phys. Rev. Lett.}\
  }\textbf {\bibinfo {volume} {104}},\ \bibinfo {pages} {240401} (\bibinfo
  {year} {2010})}\BibitemShut {NoStop}%
\bibitem [{\citenamefont {Zurek}(1981)}]{Zurek:1981cp}%
  \BibitemOpen
  \bibfield  {author} {\bibinfo {author} {\bibfnamefont {W.~H.}\ \bibnamefont
  {Zurek}},\ }\href@noop {} {\bibfield  {journal} {\bibinfo  {journal}
  {Physical Review D}\ }\textbf {\bibinfo {volume} {24}},\ \bibinfo {pages}
  {1516} (\bibinfo {year} {1981})}\BibitemShut {NoStop}%
\bibitem [{\citenamefont {Zurek}(2003{\natexlab{a}})}]{Zurek:2003fm}%
  \BibitemOpen
  \bibfield  {author} {\bibinfo {author} {\bibfnamefont {W.~H.}\ \bibnamefont
  {Zurek}},\ }\href@noop {} {\bibfield  {journal} {\bibinfo  {journal} {Reviews
  of Modern Physics}\ }\textbf {\bibinfo {volume} {75}},\ \bibinfo {pages}
  {715} (\bibinfo {year} {2003}{\natexlab{a}})}\BibitemShut {NoStop}%
\bibitem [{\citenamefont {Joos}\ \emph {et~al.}(2013)\citenamefont {Joos},
  \citenamefont {Zeh}, \citenamefont {Kiefer}, \citenamefont {Giulini},
  \citenamefont {Kupsch},\ and\ \citenamefont
  {Stamatescu}}]{joos2013decoherence}%
  \BibitemOpen
  \bibfield  {author} {\bibinfo {author} {\bibfnamefont {E.}~\bibnamefont
  {Joos}}, \bibinfo {author} {\bibfnamefont {H.~D.}\ \bibnamefont {Zeh}},
  \bibinfo {author} {\bibfnamefont {C.}~\bibnamefont {Kiefer}}, \bibinfo
  {author} {\bibfnamefont {D.~J.}\ \bibnamefont {Giulini}}, \bibinfo {author}
  {\bibfnamefont {J.}~\bibnamefont {Kupsch}}, \ and\ \bibinfo {author}
  {\bibfnamefont {I.-O.}\ \bibnamefont {Stamatescu}},\ }\href@noop {} {\emph
  {\bibinfo {title} {Decoherence and the appearance of a classical world in
  quantum theory}}}\ (\bibinfo  {publisher} {Springer Science \& Business
  Media},\ \bibinfo {year} {2013})\BibitemShut {NoStop}%
\bibitem [{\citenamefont {Wu}\ and\ \citenamefont
  {Li}(2011)}]{Wu:PhysicalReviewA:2011}%
  \BibitemOpen
  \bibfield  {author} {\bibinfo {author} {\bibfnamefont {S.}~\bibnamefont
  {Wu}}\ and\ \bibinfo {author} {\bibfnamefont {Y.}~\bibnamefont {Li}},\ }\href
  {\doibase 10.1103/PhysRevA.83.052106} {\bibfield  {journal} {\bibinfo
  {journal} {Phys. Rev. A}\ }\textbf {\bibinfo {volume} {83}},\ \bibinfo
  {pages} {052106} (\bibinfo {year} {2011})}\BibitemShut {NoStop}%
\bibitem [{\citenamefont {Wu}\ and\ \citenamefont
  {Molmer}(2009)}]{Wu:PhysicsLettersA:2009}%
  \BibitemOpen
  \bibfield  {author} {\bibinfo {author} {\bibfnamefont {S.}~\bibnamefont
  {Wu}}\ and\ \bibinfo {author} {\bibfnamefont {K.}~\bibnamefont {Molmer}},\
  }\href {\doibase 10.1016/j.physleta.2009.10.026} {\bibfield  {journal}
  {\bibinfo  {journal} {Phys. Lett. A}\ }\textbf {\bibinfo {volume} {374}},\
  \bibinfo {pages} {34} (\bibinfo {year} {2009})}\BibitemShut {NoStop}%
\bibitem [{\citenamefont {Wiseman}(2002)}]{Wiseman:2002jt}%
  \BibitemOpen
  \bibfield  {author} {\bibinfo {author} {\bibfnamefont {H.~M.}\ \bibnamefont
  {Wiseman}},\ }\href@noop {} {\bibfield  {journal} {\bibinfo  {journal}
  {Physical Review A}\ }\textbf {\bibinfo {volume} {65}},\ \bibinfo {pages}
  {032111} (\bibinfo {year} {2002})}\BibitemShut {NoStop}%
\bibitem [{\citenamefont {Horodecki}\ and\ \citenamefont
  {Horodecki}(1996)}]{Horodecki:1996fc}%
  \BibitemOpen
  \bibfield  {author} {\bibinfo {author} {\bibfnamefont {R.}~\bibnamefont
  {Horodecki}}\ and\ \bibinfo {author} {\bibfnamefont {M.}~\bibnamefont
  {Horodecki}},\ }\href@noop {} {\bibfield  {journal} {\bibinfo  {journal}
  {Physical Review A}\ }\textbf {\bibinfo {volume} {54}},\ \bibinfo {pages}
  {1838} (\bibinfo {year} {1996})}\BibitemShut {NoStop}%
\bibitem [{\citenamefont {Horodecki}\ \emph {et~al.}(2009)\citenamefont
  {Horodecki}, \citenamefont {Horodecki}, \citenamefont {Horodecki},\ and\
  \citenamefont {Horodecki}}]{Horodecki:2009gb}%
  \BibitemOpen
  \bibfield  {author} {\bibinfo {author} {\bibfnamefont {R.}~\bibnamefont
  {Horodecki}}, \bibinfo {author} {\bibfnamefont {P.}~\bibnamefont
  {Horodecki}}, \bibinfo {author} {\bibfnamefont {M.}~\bibnamefont
  {Horodecki}}, \ and\ \bibinfo {author} {\bibfnamefont {K.}~\bibnamefont
  {Horodecki}},\ }\href@noop {} {\bibfield  {journal} {\bibinfo  {journal}
  {Reviews of Modern Physics}\ }\textbf {\bibinfo {volume} {81}},\ \bibinfo
  {pages} {865} (\bibinfo {year} {2009})}\BibitemShut {NoStop}%
\bibitem [{\citenamefont {Lang}\ and\ \citenamefont
  {Caves}(2010)}]{Lang:PhysRevLett:2010}%
  \BibitemOpen
  \bibfield  {author} {\bibinfo {author} {\bibfnamefont {M.~D.}\ \bibnamefont
  {Lang}}\ and\ \bibinfo {author} {\bibfnamefont {C.~M.}\ \bibnamefont
  {Caves}},\ }\href@noop {} {\bibfield  {journal} {\bibinfo  {journal} {Phys.
  Rev. Lett.}\ }\textbf {\bibinfo {volume} {105}},\ \bibinfo {pages} {150501}
  (\bibinfo {year} {2010})}\BibitemShut {NoStop}%
\bibitem [{\citenamefont {Knill}\ and\ \citenamefont
  {Laflamme}(1998)}]{Knill:PhysRevLett:1998}%
  \BibitemOpen
  \bibfield  {author} {\bibinfo {author} {\bibfnamefont {E.}~\bibnamefont
  {Knill}}\ and\ \bibinfo {author} {\bibfnamefont {R.}~\bibnamefont
  {Laflamme}},\ }\href {\doibase 10.1103/PhysRevLett.81.5672} {\bibfield
  {journal} {\bibinfo  {journal} {Phys. Rev. Lett.}\ }\textbf {\bibinfo
  {volume} {81}},\ \bibinfo {pages} {5672} (\bibinfo {year}
  {1998})}\BibitemShut {NoStop}%
\bibitem [{\citenamefont {Zurek}(2003{\natexlab{b}})}]{Zurek:2003el}%
  \BibitemOpen
  \bibfield  {author} {\bibinfo {author} {\bibfnamefont {W.~H.}\ \bibnamefont
  {Zurek}},\ }\href@noop {} {\bibfield  {journal} {\bibinfo  {journal}
  {Physical Review A}\ }\textbf {\bibinfo {volume} {67}},\ \bibinfo {pages}
  {012320} (\bibinfo {year} {2003}{\natexlab{b}})}\BibitemShut {NoStop}%
\bibitem [{\citenamefont {Landauer}(1961)}]{Landauer:1961kg}%
  \BibitemOpen
  \bibfield  {author} {\bibinfo {author} {\bibfnamefont {R.}~\bibnamefont
  {Landauer}},\ }\href@noop {} {\bibfield  {journal} {\bibinfo  {journal} {IBM
  Journal of research and development}\ }\textbf {\bibinfo {volume} {5}},\
  \bibinfo {pages} {183} (\bibinfo {year} {1961})}\BibitemShut {NoStop}%
\bibitem [{\citenamefont {Li}\ and\ \citenamefont
  {Vitanyi}(2013)}]{li2013introduction}%
  \BibitemOpen
  \bibfield  {author} {\bibinfo {author} {\bibfnamefont {M.}~\bibnamefont
  {Li}}\ and\ \bibinfo {author} {\bibfnamefont {P.}~\bibnamefont {Vitanyi}},\
  }\href@noop {} {\emph {\bibinfo {title} {{An Introduction to Kolmogorov
  Complexity and Its Applications}}}},\ Monographs in Computer Science\
  (\bibinfo  {publisher} {Springer New York},\ \bibinfo {year}
  {2013})\BibitemShut {NoStop}%
\bibitem [{\citenamefont {Cavalcanti}\ \emph {et~al.}(2011)\citenamefont
  {Cavalcanti}, \citenamefont {Aolita}, \citenamefont {Boixo}, \citenamefont
  {Modi}, \citenamefont {Piani},\ and\ \citenamefont
  {Winter}}]{Cavalcanti:2011dy}%
  \BibitemOpen
  \bibfield  {author} {\bibinfo {author} {\bibfnamefont {D.}~\bibnamefont
  {Cavalcanti}}, \bibinfo {author} {\bibfnamefont {L.}~\bibnamefont {Aolita}},
  \bibinfo {author} {\bibfnamefont {S.}~\bibnamefont {Boixo}}, \bibinfo
  {author} {\bibfnamefont {K.}~\bibnamefont {Modi}}, \bibinfo {author}
  {\bibfnamefont {M.}~\bibnamefont {Piani}}, \ and\ \bibinfo {author}
  {\bibfnamefont {A.}~\bibnamefont {Winter}},\ }\href@noop {} {\bibfield
  {journal} {\bibinfo  {journal} {Physical Review A}\ }\textbf {\bibinfo
  {volume} {83}},\ \bibinfo {pages} {032324} (\bibinfo {year}
  {2011})}\BibitemShut {NoStop}%
\bibitem [{\citenamefont {Madhok}\ and\ \citenamefont
  {Datta}(2011)}]{Madhok:2011bh}%
  \BibitemOpen
  \bibfield  {author} {\bibinfo {author} {\bibfnamefont {V.}~\bibnamefont
  {Madhok}}\ and\ \bibinfo {author} {\bibfnamefont {A.}~\bibnamefont {Datta}},\
  }\href@noop {} {\bibfield  {journal} {\bibinfo  {journal} {Physical Review
  A}\ }\textbf {\bibinfo {volume} {83}},\ \bibinfo {pages} {032323} (\bibinfo
  {year} {2011})}\BibitemShut {NoStop}%
\end{thebibliography}%

\end{document}